\documentclass[a4paper,fleqn]{cas-dc}

\UseRawInputEncoding

\def\tsc#1{\csdef{#1}{\textsc{\lowercase{#1}}\xspace}}
\tsc{WGM}
\tsc{QE}
\tsc{EP}
\tsc{PMS}
\tsc{BEC}
\tsc{DE}

\usepackage[round, sort]{natbib}
\usepackage{graphicx}
\usepackage{float} 
\usepackage{amsmath}

\begin{document}

\pdfoutput=1

\let\WriteBookmarks\relax
\def\floatpagepagefraction{1}
\def\textpagefraction{.001}
\shorttitle{X. Zhou et al. / Journal of Biomechanic}
\shortauthors{X. Zhou et~al.}

\title [mode = title]{The Effects of Unilateral Slope Loading on Lower Limb Plantar Flexor Muscle EMG Signals in Young Healthy Males }                      
\tnotemark[1]

\tnotetext[1]{This material is supported by Department of Mechanical and Energy Engineering , Southern University of Science and Technology}

\author[1]{Xinyu Zhou}[type=editor,
                        auid = 000, bioid = 1, orcid = ****-****-****-****] \fnmark [1]
\ead{12110707@mail.sustech.edu.cn}

\credit{Conceptualization, Investigation, Methodology, Formal analysis, Project administration, Writing - Original draft}

\affiliation[1]{organization={Department of Mechanical and Energy Engineering , Southern University of Science and Technology},
                addressline={1088 Xueyuan Avenue }, 
                city={Shenzhen},
                postcode={518055}, 
                country={China}}

\author[2]{Gengshang Dong}[%
   role=,
   ]
\cormark[1]
\ead{donggs2021@mail.sustech.edu.cn}

\credit{Investigation, Formal analysis, Visualization, Writing - review \& editing}

\affiliation[2]{organization={Department of Statistics and Data Science , Southern University of Science and Technology},
                addressline={1088 Xueyuan Avenue }, 
                city={Shenzhen},
                postcode={518055}, 
                country={China}}

\author[1]{Pengxuan Zhang}[%
   role=,
   ]
\cormark[1]
\ead{12110733@mail.sustech.edu.cn}

\credit{Investigation, Experiment, Visualization}

\author[1]{Chenglong Fu}
\cormark[2]
\ead{fucl@sustech.edu.cn }
\credit{Funding acquisition, Supervision}

\fntext[fn1]{This is the first author footnote}

\author[1]{Yuquan Leng}
\cormark[2]
\ead{lengyq@sustech.edu.cn }
\credit{Supervision, Writing - review \& editing}

\fntext[fn1]{This is the first author footnote}

\cortext[cor1]{These authors should be regarded as second authors.}

\cortext[cor2]{Corresponding author}

\begin{abstract}
Different loading modes can significantly affect human gait, posture, and lower limb biomechanics. This study investigated the muscle activity intensity of the lower limb soleus muscle in the slope environment of young healthy adult male subjects under unilateral loading environment. Ten subjects held dumbbells equal to 5\% and 10\% of their body weight (BW) and walked at a fixed speed on a slope of 5\(^{\circ}\) and 10\(^{\circ}\), respectively. The changes of electromyography (EMG) of bilateral soleus muscles of the lower limbs were recorded. Experiments were performed using one-way analysis of variance (ANOVA) and multivariate analysis of variance (MANOVA) to examine the relationship between load weight, slope angle, and muscle activity intensity. The data provided by this research can help to promote the development of the field of lower limb assist exoskeleton. The research results fill the missing data when loading on the slope side, provide data support for future assistance systems, and promote the formation of relevant data sets, so as to improve the terrain recognition ability and the movement ability of the device wearer.

\end{abstract}




\maketitle

\section{Research Background}

Studies have found that different loading methods, such as different backpack carrying methods, can significantly affect human gait, posture, trunk and lower limb biomechanics, and muscle activity\citep{r1, r2}. At the same time, unilateral weight-bearing walking will increase the curvature of the body parts to the non-weight-bearing side\citep{r3, r4}, and increase the tilt and rotation of the pelvis to the non-weight-bearing side\citep{r5}. Previous studies on the electrical signals of lower limb muscles have shown that the gastrocnemius and soleus muscles play a major role in ankle plantar flexion movement, and theoretically these two muscles can produce 80\% of the total plantar flexion moment\citep{r6}. Through real-time acquisition and analysis of the muscle electrical signals of the lower limb extension and flexion muscles, the actual effect of the lower limb assist exoskeleton can be effectively measured. However, there are few studies on lower limb muscle electrical signals during unilateral weight-bearing ramp walking, and the analysis of muscle responses to complex terrain is insufficient, and the support of relevant data is lacking. The results of this experiment will help to establish a database of lower limb muscle electrical signals in complex terrain, improve the accuracy of system recognition, and provide support for more intelligent and humanized lower limb assistance exoskeleton and rehabilitation robots in the future.

At the same time, the global population aging is not optimistic. The proportion of the global population aged 65 years and older is expected to rise from 10\% in 2022 to 16\% in 2050, according to the 2022 World Population Projections Report of the Population Division of the Department of Economic and Social Affairs of the United Nations Secretariat\citep{r7}. In addition to the aging population, the number of people with disabilities and chronic diseases is also increasing year by year. According to the Lancet Medical journal in 2021, stroke and low back pain ranked 4th and 9th in the disability-adjustedlife year (DALY) index, respectively\citep{r8}. Stroke patients generally have residual motor dysfunction problems, such as asymmetric gait, weakened plantar dorsiflexion ability of ankle joint, and reduced stability of knee joint support\citep{r9}. Osteoarthritis is a growing but unresolved problem. In 2019, an estimated 528 million people were living with osteoarthritis globally, an increase of 113\% since 1990\citep{r10}. Long-term or high-intensity unilateral loading can lead to increased discomfort in the shoulder, neck and spine of the body parts\citep{r4, r5, r11}, and greater damage to the various joints of the legs, especially the frontal surface of the knee joints\citep{r12}, which is more likely to cause human injury and even increase the risk of chronic diseases such as arthritis\citep{r1, r3, r12}.

For the elderly with impaired motor function and stroke patients, if scientific and effective rehabilitation training is not given in time, it may lead to permanent loss of lower limb motor function. Therefore, it is of significant research value to explore the electrical signals of soleus muscle during unilateral weight-bearing, especially during slant walking under unilateral weight-bearing condition, which has not been studied previously. Understanding the data collection and analysis of the unilateral weight-bearing slope walking test will help to understand the potential risks of lower limb joint injury in different exercise modes, and provide data guidance for targeted prevention and treatment in the future.

\section{Methodology}

\subsection{Participants}

 Thirteen healthy adult males (age \(20.8 \pm 1.6 years\), height \(179.0 \pm 4.4 cm\), weight \(74.3 \pm 6.6 kg\)) were recruited from the university population. None had a history of musculoskeletal diseases or neurological disorders that could affect gait. Handedness was determined using the Edinburgh Handedness Inventory (Oldfield, 1971), and all 13 participants were right-handed. Prior to the experiment, each participant was given a comprehensive explanation of the experimental process, informed of any potential risks, and asked to verbally confirm their informed consent.

\subsection{Experimental Equipment}

 The experiment was conducted at the Guangdong Provincial Key Laboratory of Human-Augmentation and Rehabilitation Robotics in Universities, China.

 \subsubsection*{Instrumented Treadmill:}

To accurately capture human motion posture, an instrumented treadmill (BERTEC) capable of 0-15\(^{\circ}\) inclines and allows independent control on both sides.

\subsubsection*{Surface Electromyography (EMG) Measurement Equipment:}

Surface EMG signals are part of the signals generated by muscle contraction and can be detected by surface EMG sensors. The surface electromyography acquisition device (Delsys, Trigno Avanti Sensor) used in this project is shown in Figure \ref{FIG:1}, and the sampling frequency is 1926Hz. 


\subsubsection*{Loading:}

Standard weight dumbbell plates (0.5 kg, 1 kg, 2 kg, 2.5 kg, 3 kg, etc.) were used for loading. The body weight (BW) of each participant was calculated, and weights equivalent to 5\% and 10\% of their BW were adjusted accordingly with the dumbbell plates. Anti-slip wristbands were used to prevent the dumbbells from falling during the experiment, thereby increasing safety.

\subsection{Experimental Procedure and Data Collection}

 The purpose of this study was to investigate the effect of unilateral loading on the EMG of the extension and flexor muscles of the lower limbs in healthy young people. Studies have found that plantar flexors include plantar muscle, gastrocnemius muscle and soleus muscle at the back of human leg. During plantar flexion movement of ankle joint, the gastrocnemius muscle and soleus muscle play a major role, and theoretically these two muscles can produce 80\% of the total plantar flexion moment\citep{r6}. Studies have also shown that the soleus muscle activation is quite different under different assist parameters of the lower limb assist exoskeleton system\citep{r13}. Therefore, the use of surface electromyography (EMG) sensor to detect the degree of soleus muscle activation can effectively evaluate the effect of rehabilitation exoskeleton system, or real-time terrain recognition according to EMG. Thus, EMG of soleus was mainly researched. 

The location of the sensor was shown in Figure \ref{FIG:1}.

  During the experiment, the surface EMG sensor is attached to the skin of the target muscle surface, and the surface EMG signal is received by the sensor base station, which transmits the EMG data to the host computer through serial communication.

 To ensure the quality and stability of the surface EMG signals, the following preparations were made before sensor installation:

 \begin{enumerate}[(i)]
     \item The skin hair at the sensor attachment site was shaved.
     \item The skin and sensor surface were disinfected with alcohol wipes and allowed to air dry.
     \item The sensor was attached to the skin using specialized double-sided adhesive and secured with elastic bandages to prevent loosening or detachment during movement.
 \end{enumerate}

\begin{figure}
	\centering
	\includegraphics[width=.9\columnwidth]{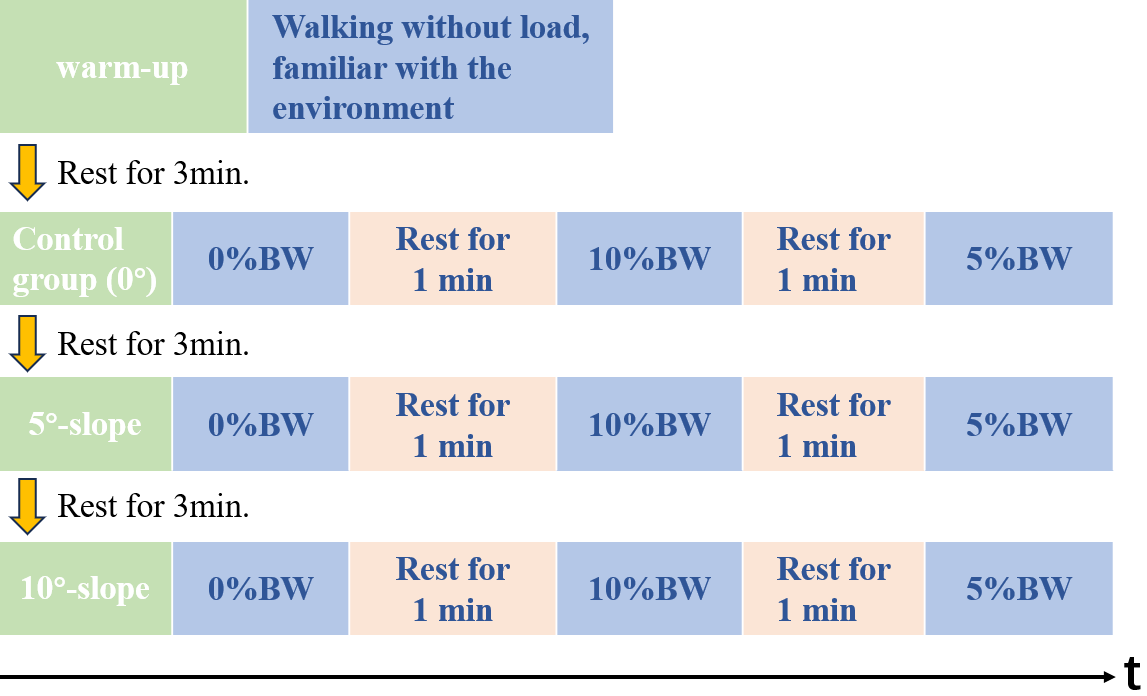}
	\caption{Experimental process}
	\label{FIG:2}
\end{figure}

 Before the formal experiment, the experimental environment was set up, and participants performed warm-up exercises to ensure the reliability of the results. The specific process was as follows: Participants walked on the treadmill without load for 3 minutes to complete the warm-up. The formal experiment began with baseline measurements on a horizontal treadmill, where participants walked for 2 minutes each without load, with a 10\% BW load, and with a 5\% BW load. This was followed by tests on 5\(^{\circ}\) and 10\(^{\circ}\) inclines, with the same loading conditions as the baseline. Sufficient rest time was provided between each test to prevent muscle fatigue, as shown in Figure \ref{FIG:2}.

\begin{figure}[h]
	\centering
	\includegraphics[width=.9\columnwidth]{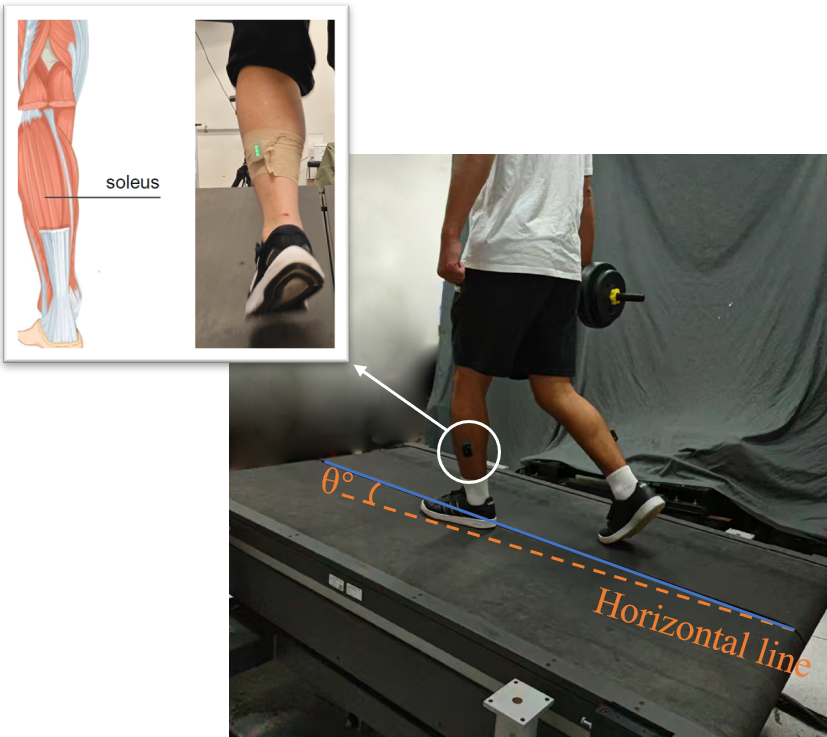}
	\caption{A subject in testing \& The surface electromyography acquisition device}
	\label{FIG:3}
\end{figure}

 The normal walking speed is approximately 1.2 m/s, but the walking speed during loading and slope walking is typically lower. Therefore, in this experiment, the speed for all test conditions was uniformly set at 0.8 m/s, as shown in Figure \ref{FIG:3}.

\subsection{Data Processing and Statistical Analysis}

 Surface EMG sensors collected EMG signals from the soleus muscle at a sampling frequency of 1926 Hz. The data were transmitted to the host computer via serial communication and processed using MATLAB 2020b. First, a 50 Hz notch filter was used to remove noise caused by 50 Hz AC power. Since surface EMG signals mainly range from 20 to 500 Hz, a second-order Butterworth bandpass filter (with cut-off frequencies of 20 Hz and 500 Hz) was used to eliminate high and low-frequency noise. Then, full-wave rectification and a second-order Butterworth low-pass filter with a cut-off frequency of 10 Hz were applied to envelope the EMG signals.

 Research indicates a relatively linear relationship between the RMS amplitude of surface EMG signals and muscle force\citep{r14}. Therefore, the RMS amplitude of surface EMG signals was used as the quantitative indicator of muscle activation. EMG signals from the soleus muscle were collected under different loading and slope conditions, and the processed RMS values were normalized(Formula (\ref{eq:standardization})) to eliminate irrelevant differences due to varying physical fitness levels among participants.

 \begin{equation}
Z = \frac{X - \mu}{\sigma}
\label{eq:standardization}
\end{equation}

\noindent where \( Z \) is the normalized value, \( X \) is the original RMS amplitude of the EMG signal, \( \mu \) is the mean RMS value, and \( \sigma \) is the standard deviation of the RMS values.\\

 The main variables in this experiment were load and treadmill slope, resulting in a total of 9 different conditions. Within each condition, the loaded side was set as the baseline value (100\%), and the relative muscle activity intensity of the non-loaded side was expressed as a percentage of this baseline. Similarly, when considering load as a single factor, the zero-load condition at the same slope was set as the baseline; when considering slope as a single factor, the zero-slope condition at the same load was set as the baseline.

 One-way Analysis of Variance (ANOVA) and Multivariate Analysis of Variance (MANOVA) were used to test for significant differences in the surface EMG signals of the soleus muscle under different conditions, with the significance level set at \(\alpha < 0.05\). ANOVA was used to analyze the significance of differences in muscle activation intensity between the two sides under 9 different conditions. MANOVA was used to analyze the effects of the three factors (load, slope, and side) on muscle activation intensity.

\section{Results}

All measurements and results in this experiment are presented as mean values with standard errors (SE), and all muscle activity intensities refer to the activity of the soleus muscle. In this study, the right side is defined as the loaded side (as all subjects were right-handed), and the left side is defined as the non-loaded side. The results obtained using the normalization method described earlier are shown in Table \ref{tbl1}, Table \ref{tbl2} and Table \ref{tbl3}.

\begin{table}[h]
\caption{0 load}\label{tbl1}
\scriptsize
\begin{tabular*}{\linewidth}{@{} LLLLLL@{} }
\toprule
Condition & Source & Sum\_square & df & F & Pr(\textgreater{}F) \\
\midrule
0\%bw 0dg & Between Groups & 0.005980 & 1 & 1.465695 & 0.242 \\
 & Within Groups & 0.073438 & 18 & & \\
\addlinespace
0\%bw 5dg & Between Groups & 0.024554 & 1 & 2.080571 & 0.166 \\
 & Within Groups & 0.212432 & 18 & & \\
\addlinespace
0\%bw 10dg & Between Groups & 0.080105 & 1 & 2.549087 & 0.128 \\
 & Within Groups & 0.565648 & 18 & & \\
\bottomrule
\end{tabular*}
\par{\footnotesize * indicates p-value \textless 0.05 (95\% confidence level)}
\end{table}

\begin{table}[h]
\caption{5\% body weight load}\label{tbl2}
\scriptsize
\begin{tabular*}{\linewidth}{@{} LLLLLL@{} }
\toprule
Condition & Source & Sum\_square & df & F & Pr(\textgreater{}F) \\
\midrule
\addlinespace
5\%bw 0dg & Between Groups & 0.099829 & 1 & 20.229065 & 0.000* \\
 & Within Groups & 0.088829 & 18 & & \\
\addlinespace
5\%bw 5dg & Between Groups & 0.061462 & 1 & 5.110177 & 0.036* \\
 & Within Groups & 0.216491 & 18 & & \\
\addlinespace
5\%bw 10dg & Between Groups & 0.034375 & 1 & 1.433766 & 0.247 \\
 & Within Groups & 0.431552 & 18 & & \\
\bottomrule
\end{tabular*}
\par{\footnotesize * indicates p-value \textless 0.05 (95\% confidence level)}
\end{table}

\begin{table}[h]
\caption{10\% body weight load}\label{tbl3}
\scriptsize
\begin{tabular*}{\linewidth}{@{} LLLLLL@{} }
\toprule
Condition & Source & Sum\_square & df & F & Pr(\textgreater{}F) \\
\midrule
\addlinespace
10\%bw 0dg & Between Groups & 0.201214 & 1 & 6.354937 & 0.021* \\
 & Within Groups & 0.569926 & 18 & & \\
\addlinespace
10\%bw 5dg & Between Groups & 0.044845 & 1 & 4.05216 & 0.059 \\
 & Within Groups & 0.199203 & 18 & & \\
\addlinespace
10\%bw 10dg & Between Groups & 0.064368 & 1 & 1.850466 & 0.191 \\
 & Within Groups & 0.626129 & 18 & & \\
\bottomrule
\end{tabular*}
\par{\footnotesize * indicates p-value \textless 0.05 (95\% confidence level)}
\end{table}

 The results indicate that in the 9 different conditions, the muscle activity intensity of the non-loaded side was significantly higher than that of the loaded side at slopes of 0\(^{\circ}\) and 5\(^{\circ}\) (p<0.05); however, this significance disappeared at a slope of 10\(^{\circ}\) (p>0.05).

\begin{figure}[h]
	\centering
	\includegraphics[width=.8\columnwidth]{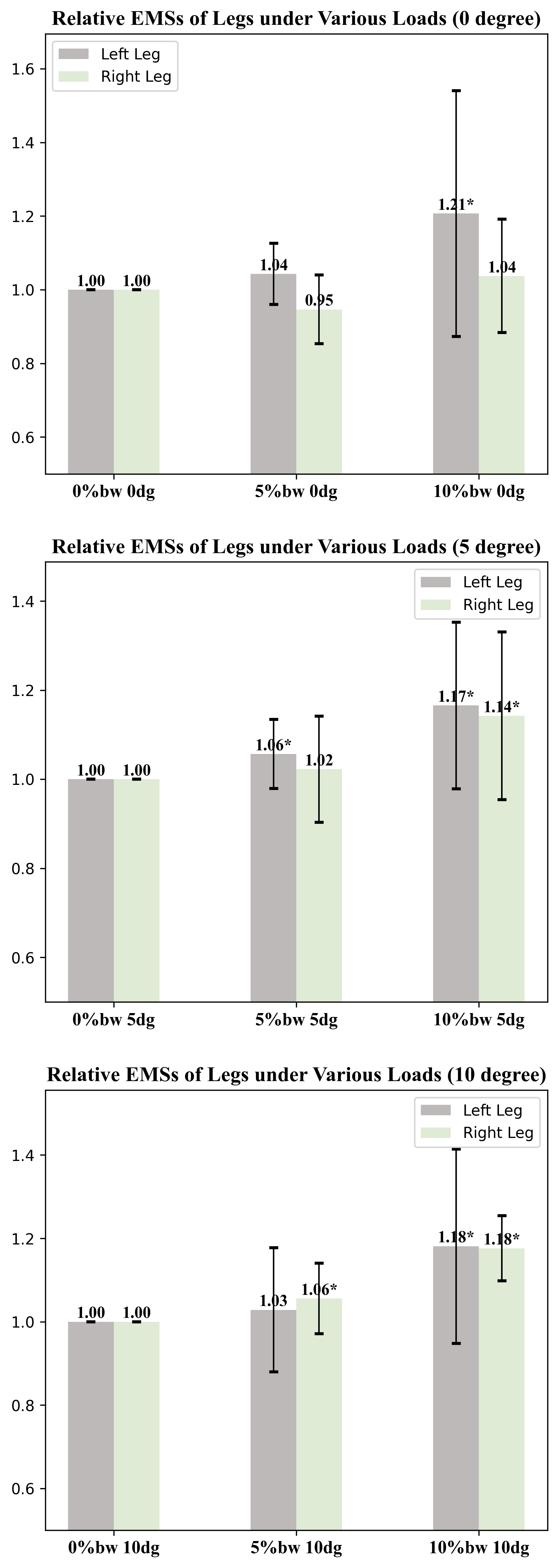}
	\caption{Relative EMSs of Legs on Various Slopes}
	\label{FIG:4}
\end{figure}

\begin{figure}[h]
	\centering
	\includegraphics[width=.8\columnwidth]{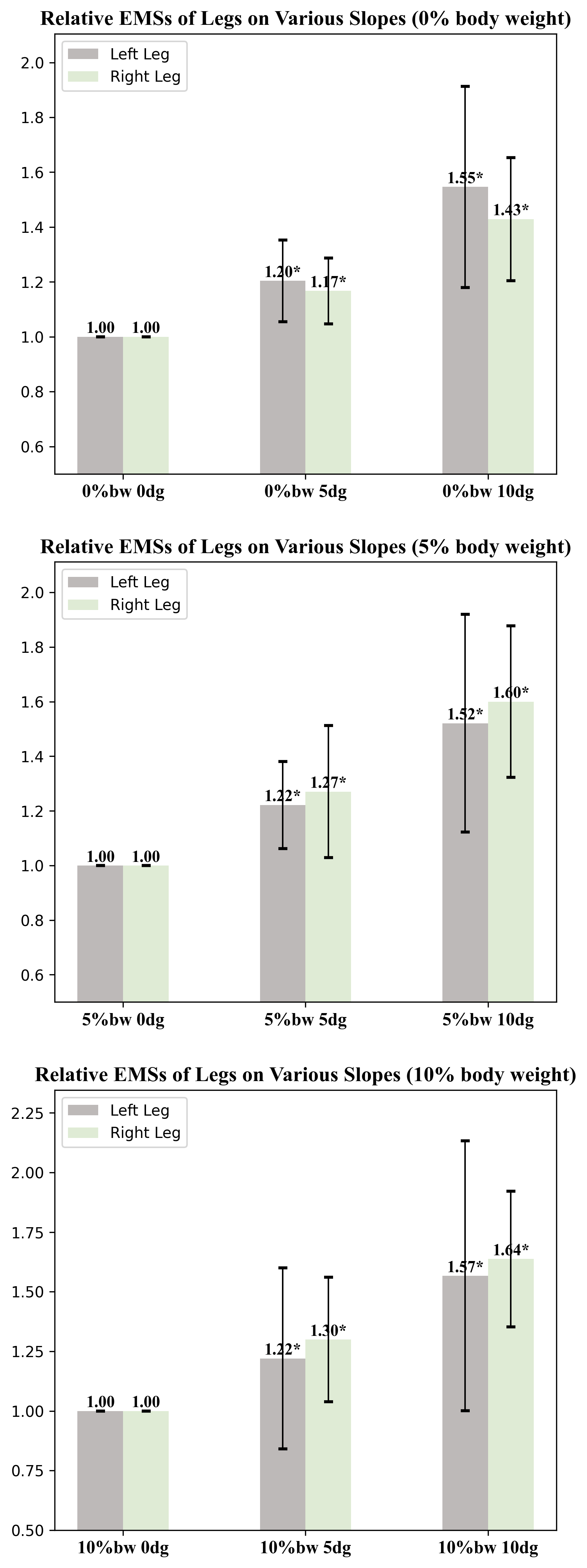}
	\caption{Relative EMSs of Legs under Various Loads}
	\label{FIG:5}
\end{figure}

 The integrated electromyography (iEMG) results are shown in Figure \ref{FIG:4} and Figure \ref{FIG:5}. 

 When analyzing the load as a single factor, it was found that at a treadmill slope of 0\(^{\circ}\), the muscle activity intensity at 5\%BW load was significantly higher than that at 0\%BW load, and the muscle activity intensity at 10\%BW load was significantly higher than that at 5\%BW load. At a slope of 5\(^{\circ}\), the muscle activity intensity at 5\%BW was significantly higher than that at 0\%BW, and the muscle activity intensity at 10\%BW was significantly higher than that at 5\%BW. At a slope of 10\(^{\circ}\), the muscle activity intensity at 5\%BW was significantly higher than that at 0\%BW, and the muscle activity intensity at 10\%BW was significantly higher than that at 5\%BW.

 When considering slope as a single factor, it was found that at 0\%BW load, the muscle activity intensity at a slope of 5\(^{\circ}\) was significantly higher than that at 0\(^{\circ}\), and the muscle activity intensity at a slope of 10\(^{\circ}\) was significantly higher than that at 5\(^{\circ}\). This significance also held true for 5\%BW and 10\%BW loads.

 The results of the multivariate analysis of variance (MANOVA) are shown in Table \ref{tbl4}. 

 \begin{table}[h]
\caption{Statistical Results of MANOVA}\label{tbl4}
\scriptsize
\begin{tabular*}{\linewidth}{@{} LLLLLL@{} }
\toprule
Source & {Sum\_square} & {df} & {F} & {Pr(\textgreater{}F)} \\
\midrule
C(Side) & 3636.70 & 1 & 6.45 & 0.012* \\
C(Weight) & 58231.39 & 2 & 51.61 & 0.000* \\
C(Degree) & 7566.73 & 2 & 6.71 & 0.002* \\
C(Side):C(Weight) & 47.19 & 2 & 0.04 & 0.959 \\
C(Side):C(Degree) & 350.36 & 2 & 0.31 & 0.733 \\
C(Weight):C(Degree) & 682.48 & 4 & 0.30 & 0.876 \\
C(Side):C(Weight):C(Degree) & 305.45 & 4 & 0.14 & 0.969 \\
Residual & 91385.60 & 162  \\
\bottomrule
\end{tabular*}
\par{\footnotesize * indicates p-value \textless 0.05 (95\% confidence level)}
\end{table}

 It is clear that slope (Degree) has the most significant effect on EMG signals, followed by load (Weight). The left and right legs (Group) also have significant effects, but they are relatively smaller. From the MANOVA, all considered interactions were not significant, indicating that the effects of different factor combinations on the results can be considered independent. This means that the effects of slope, load, and the distinction between the left and right legs mainly occur independently without significant interaction effects impacting the strength of the EMG signals.

\section{Analysis}

Existing research on unilateral loading primarily focuses on the impact of load on trunk and lower limb muscle activity during walking, isolated from EMG of soleus as well as studies on the effects of slopes on the plantar flexor muscle groups. Therefore, this study aims to be a supplement the previous experiments and combine the factors of slope and unilateral loading, discussing their potential relationship based on previous research conclusions.

\subsection{Plantar Flexor Muscle iEMG Analysis}

 In this study, when the weight was analyzed as a single factor variable, when the slope of the running platform was 0\(^{\circ}\), 5\(^{\circ}\) and 10\(^{\circ}\), the muscle activity intensity of both sides under 5\%BW weight was significantly greater than that under 0\%BW weight, and the muscle activity intensity under 10\%BW weight was significantly greater than that under 5\%BW weight. When slope is considered as a single factor variable, it is found that the muscle activity intensity at a slope of 5\(^{\circ}\) is significantly greater than that at a slope of 0\(^{\circ}\), and that at a slope of 10\(^{\circ}\) is significantly greater than 5\(^{\circ}\). This significance held true at both 5\%BW and 10\% load.

According to previous studies, the presence of external loads and the increase in weight as well as the slope of the ground (treadmill) have been shown to affect the activity of key muscles involved in gait, thus affecting the stability of human movement\citep{r15, r16, r17, r18}. In the experiment, the soleus iEMG activity increased significantly with the increase of unilateral load and slope, indicating that the extension and flexion muscles of the lower limbs were trying to maintain balance in the face of unilateral load.

\subsection{One-way ANOVA of bilateral differences}

\subsubsection{About unilateral loading}

According to the previous literature on unilateral loading, there are significant differences between the left and right muscles, load-bearing muscles and non-load-bearing muscles. In unilateral weight-bearing examination, peak EMG of the lateral femoris, semimembranes, gluteus medius, and gastrocnemius increased\citep{r1, r19}, but the response of soleus in the extension and flexion muscles of the lower extremities to unilateral weight-bearing remains unclear.

In this experiment provided 9 different working conditions (0\(^{\circ}\), 5\(^{\circ}\), 10\(^{\circ}\); With 0\%BW0\(^{\circ}\), 0\%BW5\(^{\circ}\), and 0\%BW10\(^{\circ}\) as the basis), there was no significant difference in muscle activity intensity between the load-bearing side and the load-bearing side under the three working conditions with a load of 0kg (p>0.1), and only when there was a unilateral load, there might be a significant difference in muscle activity intensity between the heavy side and the load-bearing side. In fact, it was observed in the test that there were significant differences in muscle activity intensity between the load-bearing side and the load-bearing side (p<0.05) under the four working conditions of 5\%BW at 0\(^{\circ}\), 10\%BW at 0\(^{\circ}\), 5\%BW at 5\(^{\circ}\) and 10\%BW at 5\(^{\circ}\), and the non-load-bearing side was greater than the load-bearing side. No significant difference was found between 5\%BW at slope 10\(^{\circ}\) and 10\%BW at slope 10\(^{\circ}\), but the mean activity intensity of the non-loading side was still greater than that of the loading side.

Therefore, the experimental results when the slope was 0\(^{\circ}\) and 5\(^{\circ}\) were consistent with the previous conclusion that the unilateral load experiment led to the increase of the lower limb muscle activity intensity on the non-bearing side\citep{r1, r2, r16}. Although the dumbbell used as the load in this experiment was different from the previous load forms such as the unilateral backpack, the researchers believed that the results should still be similar.

\subsubsection{About Slope walking}

Previous experiments on slope walking have shown that the activity intensity of lower limb muscles will be strengthened with the increase of slope\citep{r18, r20}, and the muscle activity intensity of lower limb extension and flexion muscles (including soles) will be significantly improved\citep{r21}. Although there have been slope experiments on no-load and bilateral load, the results of unilateral load slope experiments are still unclear.

In the experiment, 9 working conditions were divided into three groups (0\%BW, 5\%BW, 10\%BW; They were normalized with 0\%BW0\(^{\circ}\), 5\%BW0\(^{\circ}\) and 10\%BW0\(^{\circ}\), respectively. The results showed that the results were similar to the conclusions of previous studies when no weight was loaded, and the muscle activity intensity of soles increased with the increase of slope. Similar results were also shown under the conditions of 5\%BW and 10\%BW load. Meanwhile, the researchers found that the iEMG average value of muscle activity intensity on the load-bearing side was higher under the condition of unilateral load, that is, with the increase of slope, the muscle activity intensity on the load-bearing side increased more significantly than that on the non-load-bearing side.

\subsection{Multivariate analysis of variance considering load and slope}

In this experiment, 9 groups of different experimental conditions were grouped and normalized according to load and slope to obtain different factors. Meanwhile, multivariate analysis of variance was used to try to combine unilateral load with experimental characteristics of slope walking, and a more comprehensive conclusion was obtained.

From the multivariate analysis of variance (Table 3), C(Side):C(Weight):C(Degree)>0.9 indicates that the interaction of the three factors (load bearing, slope, load bearing side and non-load bearing side) is not significant, indicating that the influence of different factor combinations on the results can be considered independent. That is, the effects of slope, load bearing side and difference between load bearing side and non-load bearing side are mainly independent, and there is no obvious interaction effect on muscle activity intensity.

Experimental data show (FIG. 2) that unilateral load can improve the muscle activity intensity of the non-load bearing side, and the muscle activity intensity is positively correlated with the load. Slope walking can significantly increase the muscle activity intensity of the extension and flexion muscles of lower limbs, and the muscle activity intensity is positively correlated with slope.

From the perspective of the significance of the difference between the load-bearing side and the non-load-bearing side, the reduction of significance (Table 1) is reasonable, because it is observed in the iEMG figure (Figure 2) that the mean value of iEMG of muscle activity intensity on the load-bearing side is higher when unilateral load is present, that is, with the increase of slope, the increase of muscle activity intensity on the load-bearing side is more significant than that on the non-load-bearing side.

Slope (Degree) has the most significant effect on EMG, followed by Weight (Weight), and left and right legs (Group) also have a significant effect, but relatively small.

\subsection{Limitations}

 This study has several limitations, including the inability to accurately quantify the actual effect of using dumbbells as the ideal load. The swing of the center of gravity when holding dumbbells is higher than when carrying bagged items, but lower than when holding items clamped under one arm. These different loading methods may affect the experiment's conclusions. Additionally, based on previous experimental conclusions, individual differences in actions and postures when carrying loads were not distinguished. These primarily fall into two categories: bending the trunk towards the load and bending towards the non-load side. This study did not specifically differentiate these two scenarios, but it appears that these different bending methods did not significantly affect the results.

 Furthermore, unavoidable factors such as the participants' muscle distribution and the lack of more sophisticated measurement tools could cause slight changes in the position of the EMG sensors on the skin, affecting the recorded activity. There may also be crosstalk from the activity of adjacent muscles. 

 The sample selection also has room for improvement. The current sample size is somewhat insufficient, and the participants were limited to young adult males.

\section{Conclusion}
This study supplements previous experiments focusing on the activity of lower limb plantar flexor muscles under unilateral loading. It validates several conclusions from earlier studies, such as the increase in muscle activity of the lower limb plantar flexor muscles with increased load weight, and the observation that the non-loaded side (left side) exhibits higher muscle activity compared to the loaded side (right side). Additionally, this study fills a gap in research concerning EMG signals of the lower limb plantar flexor muscles under slope conditions with unilateral loading.

 The research indicates that at slopes of 0\(^{\circ}\) and 5\(^{\circ}\), the muscle activity intensity of the non-loaded side is significantly higher than that of the loaded side. However, this significance disappears at a slope of 10\(^{\circ}\), though the average muscle activity of the non-loaded side still exceeds that of the loaded side. Therefore, it can be inferred that at lower slopes, increasing the load weight will lead to increased muscle activity in the lower limb plantar flexor muscles, with the non-loaded side showing higher activity than the loaded side. These conclusions, based on ground level tests, remain valid, but as the slope increases, the impact of unilateral loading decreases significantly.

 The study also found that under the influence of slope as a single variable, regardless of load weight, the growth rate of muscle activity intensity on the loaded side is significantly greater than that on the non-loaded side. This result is reasonable because at lower slopes, the muscle activity intensity on the loaded side is significantly lower than that on the non-loaded side, whereas at higher slopes (10\(^{\circ}\) in this experiment), the difference between the two sides is not obvious. This is because the slope factor reduces the impact of unilateral loading.

 It is recommended that individuals, especially the elderly, alternate sides when carrying loads unilaterally to avoid asymmetric muscle fatigue. If possible, unilateral loads should be limited, preferably not exceeding 5\% BW. Additionally, it is noteworthy that walking on slopes with a load can reduce the negative effects of unilateral loading. Therefore, for certain work scenarios that require carrying loads up slopes, it may be worth considering the use of unilateral loading based on actual conditions.

\printcredits



\bibliographystyle{apalike}
\bibliography{cas-refs}






\end{document}